\documentclass{jps-cp}

\usepackage{color}
\usepackage[normalem]{ulem}
\usepackage{txfonts} 

\title{
Efficiency of Continuous-Time Quantum Monte Carlo Updates in the Ferromagnetic State of the Doped SU(3) Fermi-Hubbard Model
}
\author{Juntaro \textsc{Fujii}$^{1}$, Kazuki \textsc{Yamamoto}$^{2,3,4}$ and Akihisa \textsc{Koga}$^{1,5}$}

\inst{$^{1}$Department of Physics, Institute of Science Tokyo, Meguro, Tokyo 152-8551, Japan \\
$^{2}$Research Institute for Innovation and Co-Creation, Osaka Metropolitan University, Sakai, Osaka 599-8531, Japan \\
$^{3}$Department of Physics, Osaka Metropolitan University, Sumiyoshi, Osaka 558-8585, Japan \\
$^{4}$Nambu Yoichiro Institute of Theoretical and Experimental Physics (NITEP),
Osaka Metropolitan University, Sumiyoshi, Osaka 558-8585, Japan\\
$^{5}$Department of Physics, Chuo University, Bunkyo, Tokyo 112-8551, Japan}

\email{fujii.j.47dc@m.isct.ac.jp}

\recdate{September 20, 2026}

\abst{
We investigate the sampling efficiency
in a segment-based continuous-time hybridization-expansion quantum Monte Carlo solver 
within dynamical mean-field theory for the doped SU(3) Fermi-Hubbard model.
Focusing on the low-temperature and strong-coupling regime, 
where a ferromagnetic state appears upon hole doping away from one-third filling, 
we compare several update schemes.
Adding the double-flip update to the basic local update yields the smallest integrated autocorrelation time 
of the majority-flavor occupation in the ferromagnetic regime, 
whereas adding the flavor-permutation update is the most effective in the paramagnetic regime.  
We further clarify the origin of this difference 
by analyzing the characteristic occupation changes 
caused by accepted updates.
}

\kword{SU($N$) Fermi-Hubbard model, Ferromagnetism, CT-HYB}

\begin{document}
\maketitle
\section{Introduction}
Multicomponent fermionic systems with SU($N$) symmetry 
have attracted considerable interest 
as a platform for studying correlation effects 
beyond the conventional SU(2) case~\cite{CazalillaRey2014,IbarraGarciaPadillaChoudhury2025,OkanamiTakemoriKoga2014,YanatoriKoga2016,KogaYanatori2017}.
In particular, the SU($N$) Fermi-Hubbard model provides
a setting for studying itinerant ferromagnetism
in systems with enlarged internal symmetry~\cite{FujiiYamamotoKoga2026,JinNie2026,TamuraKatsura2023}.
In the limit of infinite interaction strength with a single hole, 
a fully polarized ferromagnetic ground state has been established
for the SU($N$) Fermi-Hubbard model~\cite{KatsuraTanaka2013,BotzungNataf2024}.
More recently, calculations based on dynamical mean-field theory (DMFT) for the doped SU($3$) Fermi-Hubbard model
have shown that ferromagnetic (FM) order appears at low temperatures in the strong-coupling regime,
and that the transition between the FM and paramagnetic (PM) states is of first order~\cite{FujiiYamamotoKoga2025}.

In DMFT calculations, the effective impurity problem is commonly solved 
using the continuous-time hybridization-expansion quantum Monte Carlo method (CT-HYB)~\cite{WernerEtAl2006,GullEtAl2011}.
In the segment representation, however, Monte Carlo sampling can become inefficient
in the strong-coupling regime, particularly near commensurate filling.
This is because conventional local updates,
such as the insertion and removal of segments and antisegments,
can have low acceptance probabilities. 
To improve the sampling efficiency in this regime, several additional updates have been introduced, 
including global flavor-permutation and double-flip updates~\cite{GullEtAl2011,KogaWerner2011}. 
In particular, the efficiency of double-flip updates 
has been discussed in the SU($2$) case~\cite{KogaKamogawaNasu2021}.
In contrast, how the choice of update scheme affects the autocorrelation of physical observables, 
particularly in the doped metallic regime, has not been well clarified.

In this paper, we compare several Monte Carlo update schemes
for the doped SU($3$) Fermi-Hubbard model at low temperatures in the strong-coupling regime.
We focus on a parameter region where both FM and PM solutions exist
and evaluate the integrated autocorrelation time of the majority-flavor occupation
together with the acceptance ratios. 
We show that, among the update schemes considered, 
adding the double-flip update yields the smallest integrated autocorrelation time of the majority-flavor occupation in the FM state, 
whereas adding the flavor-permutation update yields the smallest integrated autocorrelation time in the PM state. 
To identify the configuration changes associated with this improvement,
we compare the occupation distributions with the changes caused by accepted updates.
\section{Model and Method}
We consider the SU($3$) Fermi-Hubbard model on a hypercubic lattice, described by the Hamiltonian
\begin{equation}
\textstyle
H =
-t \sum_{\langle i,j\rangle,\sigma}
\left(
\hat{c}_{i\sigma}^{\dagger}\hat{c}_{j\sigma}+\mathrm{H.c.}
\right)
+
\frac{U}{2}
\sum_{i,\sigma\neq\sigma'}
\hat{n}_{i\sigma}\hat{n}_{i\sigma'},
\end{equation}
where $\hat{c}_{i\sigma}^{\dagger}$ ($\hat{c}_{i\sigma}$) creates (annihilates) a fermion 
with flavor $\sigma=0,1,2$ at site $i$, 
and $\hat{n}_{i\sigma}=\hat{c}_{i\sigma}^{\dagger}\hat{c}_{i\sigma}$. 
Here, $t$ is the nearest-neighbor hopping amplitude 
and $U$ is the repulsive on-site interaction. 
To keep the kinetic energy finite in the infinite-dimensional limit, the hopping amplitude is scaled as $t \propto d^{-1/2}$.

We study this model using DMFT~\cite{MetznerVollhardt1989,GeorgesEtAl1996}.
In the infinite-dimensional limit, the self-energy becomes local and momentum independent, 
$\Sigma_{\sigma}(\mathbf{k},i\omega_n)=\Sigma_{\sigma}(i\omega_n)$,
where $\omega_n=(2n+1)\pi T$ is the fermionic Matsubara frequency.
The local Green's function is then given by
\begin{equation}
G_{\sigma}(i\omega_n)
=
\int d\epsilon\,
\frac{\rho_0(\epsilon)}
{i\omega_n+\mu-\epsilon-\Sigma_{\sigma}(i\omega_n)},
\end{equation}
where $\mu$ is the chemical potential and controls the filling. 
For the infinite-dimensional hypercubic lattice, the noninteracting density of states is given by 
$\rho_0(\epsilon)=(\sqrt{\pi}D)^{-1}\exp[-(\epsilon/D)^2]$.
We use $D$ as the energy unit. 

Within DMFT, the lattice problem is mapped onto an effective impurity problem, for which the Dyson equation is written as
$\mathcal{G}_{\sigma} (i\omega_n)^{-1} = G_{\mathrm{imp}, \sigma} (i\omega_n)^{-1} + \Sigma_{\mathrm{imp},\sigma}(i\omega_n)$,
where $\mathcal{G}_{\sigma} (i\omega_n)$ is the Weiss Green's function. 
The lattice self-energy is identified with the impurity self-energy,
$\Sigma_{\sigma}(i\omega_n) = \Sigma_{\mathrm{imp}, \sigma} (i\omega_n)$,
and the impurity problem is solved self-consistently to satisfy $G_{\sigma}(i\omega_n) = G_{\mathrm{imp},\sigma}(i\omega_n)$. 
We allow flavor-dependent Green's functions and self-energies, 
so that the SU(3) flavor symmetry can be spontaneously broken.

We solve the effective impurity problem using the CT-HYB method in the segment representation 
at inverse temperature $\beta=1/T$.
In CT-HYB, Monte Carlo configurations are sampled using the Metropolis-Hastings algorithm~\cite{MetropolisEtAl1953,Hastings1970}, 
and physical observables are evaluated as Monte Carlo averages. 
For density-density interactions, a Monte Carlo configuration can be represented by a set of segments for each flavor. 
Here, we represent a configuration as $\mathcal{C}=\{\mathcal{S}_{\sigma}\}_{\sigma=0}^{N-1}$,
where $\mathcal{S}_{\sigma}$ is either 
$\varnothing$,
$[0,\beta)$,
or $\{[\tau_{\sigma m}^{\mathrm{s}},\tau_{\sigma m}^{\mathrm{e}})\}_{m=1}^{k_\sigma}$,
corresponding to an empty line,
a full line, and a set of $k_\sigma$ segments, respectively.
$\tau_{\sigma m}^{\mathrm{s}}$ ($\tau_{\sigma m}^{\mathrm{e}}$) denotes the start (end) time 
of the $m$-th segment of flavor $\sigma$. 
For each flavor, the start and end points of the segments alternate along the imaginary-time axis,
with imaginary time understood modulo $\beta$.

For a given configuration $\mathcal{C}$, 
we define the occupation of flavor $\sigma$ as 
\begin{equation}
x_\sigma
=
\begin{cases}
0, & \mathcal{S}_\sigma=\varnothing,\\
1, & \mathcal{S}_\sigma=[0,\beta),\\
\textstyle
\frac{1}{\beta}
\sum_{m=1}^{k_\sigma}
[(\tau_{\sigma m}^{\mathrm{e}}
-\tau_{\sigma m}^{\mathrm{s}})\bmod\beta],
& \text{otherwise}.
\end{cases}
\end{equation}
This quantity determines the flavor-resolved particle density 
through its Monte Carlo average as 
$n_\sigma \equiv\langle \hat{n}_{i\sigma}\rangle=\left\langle x_\sigma\right\rangle_{\mathrm{MC}}$.
The total particle density is then given by $n= \sum_\sigma n_\sigma$. 

The basic local update of the Monte Carlo configuration 
consists of the insertion or removal of a segment or an antisegment,
where an antisegment is the unoccupied interval
between two neighboring segments of the same flavor~\cite{WernerEtAl2006,GullEtAl2011}. 
In the following, we refer to this local insertion/removal update as the simple update.
To improve the sampling efficiency, we also consider a flavor-permutation update~\cite{GullEtAl2011}. 
In this update, two flavors $\sigma$ and $\sigma'$ are selected,
and their entire segment configurations 
are exchanged $(\mathcal{S}_{\sigma},\mathcal{S}_{\sigma'})\rightarrow(\mathcal{S}_{\sigma'},\mathcal{S}_{\sigma})$.

We further consider the double-flip update~\cite{KogaWerner2011}. 
To define this update, two flavors $\sigma$ and $\sigma'$ are first selected.
For the two selected flavors, all segment endpoints are arranged 
in a single sequence in imaginary-time order, with their flavor labels retained.
Two neighboring endpoints in this periodic sequence are then selected.
If the proposal is accepted, the flavor label of each endpoint is switched between $\sigma$ and $\sigma'$.

\begin{figure}[t]\begin{center}
\includegraphics[width=\textwidth]{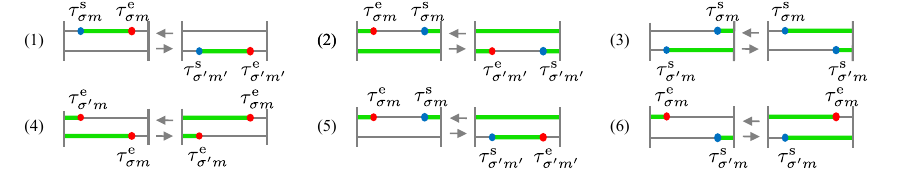}
\end{center}\caption{
Schematic illustration of the six types of double-flip updates.
The two horizontal lines represent the local imaginary-time configurations 
of the two flavors involved in the update. 
Blue (red) dots indicate segment start (end) points, 
$\tau_{\sigma m}^{\mathrm{s}}(\tau_{\sigma m}^{\mathrm{e}})$,
and green lines represent segments.
The arrows indicate the corresponding changes in the segment configurations induced by each update.
Update (1) corresponds to the type-1 double-flip update discussed in the text.
}\label{fig_1}\end{figure}

Depending on the local configuration around the selected endpoints, 
there are six possible double-flip patterns, as illustrated in Fig.~\ref{fig_1}.
In the type-1 pattern, the selected endpoints are the start and end points of a segment in one flavor, 
while the other flavor is unoccupied over the same interval.
When accepted, this update transfers the segment to the other flavor.
We note that a closely related update is implemented in the CTSEG impurity solver~\cite{KavokineEtAl2025}.
The remaining five patterns are referred to here as the other double-flip updates.

In the following, we compare three Monte Carlo update schemes: 
the simple update alone, the simple update combined with the flavor-permutation update, 
and the simple update combined with the double-flip update.
The flavors or flavor pairs involved in each update
are selected with equal probability.
Segment insertion/removal, antisegment insertion/removal,
and the additional update, when included,
are selected with equal probability.
Since the first two processes constitute the simple update,
the relative proposal frequency of the simple update
and the additional update is $2:1$.
We define one Monte Carlo step (MCS) as one proposal of any update
included in the corresponding update scheme.
Measurements are performed after every MCS.
\section{Results}
\begin{figure}[t]\begin{center}
\includegraphics[width=\textwidth]{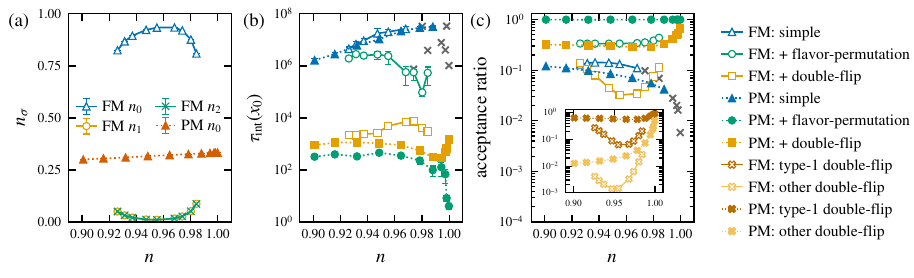}
\end{center}\caption{
(a) Flavor-resolved particle densities $n_{\sigma}$ 
for the FM and PM solutions 
as functions of the total particle density $n$. 
(b) Integrated autocorrelation time $\tau_{\mathrm{int}}(x_0)$
of the occupation $x_0$ in the FM and PM solutions
for the simple, simple $+$ flavor-permutation, and simple $+$ double-flip update schemes.
(c) Acceptance ratios of the simple, flavor-permutation, and double-flip updates.
Solid and dotted lines indicate the FM and PM solutions, respectively.
The inset shows the acceptance ratios of the type-1 and other double-flip updates separately. 
Crosses indicate data points from simple-update runs 
that did not reach equilibrium and are shown only for reference.
The results are obtained at $T/D = 0.005$ and $U/D = 300$ using $10^{10}$ MCS. 
}\label{fig_2}\end{figure}

As shown in our previous study~\cite{FujiiYamamotoKoga2025}, at low temperatures and for strong interactions, 
hole doping away from one-third filling induces a FM state
in which the particle density is strongly polarized toward a single flavor. 
Owing to the first-order transition between the FM and PM states,
both a flavor-polarized FM solution and a flavor-symmetric PM solution are obtained as self-consistent DMFT solutions over a finite doping range.
Figure~\ref{fig_2}(a) shows the flavor-resolved particle densities $n_\sigma$ 
of these two solutions at $T/D = 0.005$ and $U/D = 300$. 
In the FM solution, the density $n_0$ is much larger than $n_1$ and $n_2$, 
whereas the three densities are equal in the PM solution.

To examine the sampling efficiency of these solutions, 
we calculate the integrated autocorrelation time of the majority-flavor occupation $x_0$.
For an observable $X$, the integrated autocorrelation time is related 
to the variance of its Monte Carlo average by
\begin{equation}
\mathrm{Var}(\left\langle X\right\rangle_{\mathrm{MC}}) \simeq \frac{2 \tau_{\mathrm{int}} (X)}{N_{\mathrm{MCS}}} \mathrm{Var}(X), 
\end{equation}
where $N_{\mathrm{MCS}}$ is the total number of Monte Carlo steps 
and is assumed to be sufficiently large.
Thus, for a given observable and fixed $N_{\mathrm{MCS}}$,
a smaller $\tau_{\mathrm{int}}$ corresponds to a smaller statistical error.
We estimate $\tau_{\mathrm{int}}$ numerically
from the variance of binned data~\cite{Janke2002}.

Figure~\ref{fig_2}(b) shows $\tau_{\mathrm{int}}(x_0)$ for the simple, 
simple $+$ flavor-permutation, and simple $+$ double-flip update schemes. 
Apart from the unequilibrated points marked by crosses, the simple update alone gives very long
$\tau_{\mathrm{int}}$ for both the FM and PM solutions.
Adding a flavor-permutation or double-flip update reduces $\tau_{\mathrm{int}}$.
In the FM solution, the simple $+$ double-flip update scheme is the most effective,
whereas in the PM solution the simple $+$ flavor-permutation update scheme gives the smallest integrated autocorrelation time.

To examine whether this difference simply reflects the acceptance ratio of each update, 
we compare the corresponding acceptance ratios, shown in Fig.~\ref{fig_2}(c). 
Here, the acceptance ratio is defined 
as the number of accepted updates divided by the number of proposed updates. 
In the PM solution, the flavor-permutation update has the highest acceptance ratio, 
followed by the double-flip and simple updates. 
In contrast, in the FM solution, the acceptance ratio of the double-flip update 
is lower than those of the other two updates over most of the density range. 
Therefore, the small $\tau_{\mathrm{int}}$ obtained with the simple $+$ double-flip update scheme in the FM solution 
cannot be explained by the acceptance ratio of the double-flip update alone.

\begin{figure}[t]\begin{center}
\includegraphics[width=\textwidth]{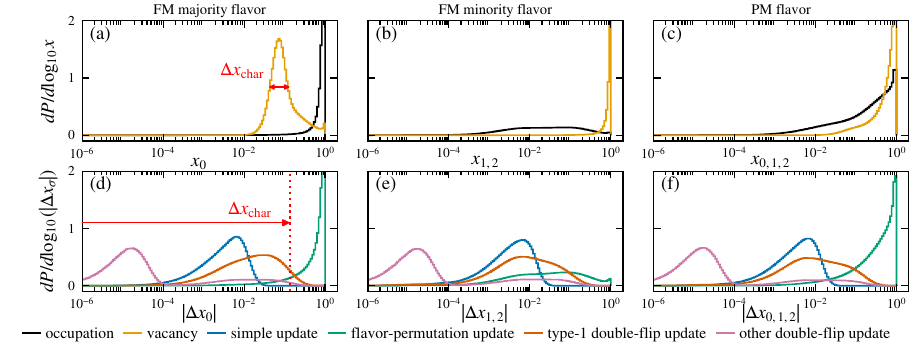}
\end{center}\caption{
Distributions of the occupation and vacancy for each flavor, 
and of occupation changes caused by accepted updates. 
(a)--(c) Probability densities $dP/d \log_{10} x$ sampled after each MCS 
for the FM majority flavor, FM minority flavor, and one flavor in the PM solution, respectively. 
Here, $x = x_\sigma$ for the occupation (black) 
and $x = 1 - x_\sigma$ for the vacancy (yellow). 
(d)--(f) Probability densities $dP/d\log_{10}(|\Delta x_{\sigma}|)$ 
of the magnitude of the occupation change of flavor $\sigma$
caused by accepted simple (blue), flavor-permutation (green), type-1 double-flip (orange), 
and other double-flip (magenta) updates.
The red arrow in (a) indicates the characteristic vacancy width 
$\Delta x_{\mathrm{char}}$, defined as the difference between the two $x$ values 
at half the maximum of the vacancy peak.
The red arrow in (d) shows the same scale on the $|\Delta x_0|$ axis, 
and the red dotted line marks the corresponding position.
The results are obtained at $T/D = 0.005$, $U/D = 300$, and $n \simeq 0.93$.
}\label{fig_3}\end{figure}

To clarify the origin of this behavior,
we compare the characteristic scales of the occupation and vacancy distributions
with those of the occupation changes caused by accepted updates
at $n \simeq 0.93$, as shown in Fig.~\ref{fig_3}.
The logarithmic probability densities represent the probability weight per logarithmic interval.
For the majority flavor in the FM solution, the vacancy distribution exhibits a pronounced peak
at a finite value, as shown in Fig.~\ref{fig_3}(a), 
which is absent for the FM minority flavor [Fig.~\ref{fig_3}(b)] 
and the PM flavor [Fig.~\ref{fig_3}(c)]. 
We characterize the width of this peak by $\Delta x_{\mathrm{char}}$,
defined as the difference between the two vacancy values at half maximum.
Interestingly, Fig.~\ref{fig_3}(d) shows that the distribution of $|\Delta x_0|$
caused by the type-1 double-flip update has substantial weight
around $\Delta x_{\mathrm{char}}$.
In contrast, the simple update has relatively little weight on this scale,
whereas the flavor-permutation update mainly produces larger changes.
Changes much smaller than $\Delta x_{\mathrm{char}}$ require many successive accepted moves
to traverse the characteristic width, whereas changes larger than $\Delta x_{\mathrm{char}}$
cannot directly connect two configurations within this characteristic interval.
Thus, the type-1 double-flip update is particularly well suited 
to changing $x_0$ on the characteristic scale of the vacancy in the FM majority flavor. 

We next compare the type-1 double-flip update 
with the simple update in more detail. 
As shown in Figs.~\ref{fig_3}(d)--\ref{fig_3}(f), 
the distribution for the type-1 double-flip update extends 
to larger $|\Delta x_{\sigma}|$ 
and has substantially more weight in this region 
than that for the simple update. 
On the other hand, when focusing on a single imaginary-time axis, 
both updates produce a change in $x_\sigma$ associated with a single segment. 
The difference in their distributions therefore indicates that 
relatively long segments are present
in the Monte Carlo configurations,
but that accepted simple updates less frequently produce occupation changes 
on the corresponding scale.
In contrast, the type-1 double-flip update can generate such occupation changes
by transferring an existing long segment between flavors,
rather than by inserting or removing it.
This is likely to contribute to the strong reduction 
of $\tau_{\mathrm{int}}(x_0)$ observed in Fig.~\ref{fig_2}(b). 

The flavor-permutation update plays a different role.
By exchanging the entire imaginary-time configurations of two flavors,
it produces an occupation change
$|\Delta x_\sigma|=|x_{\sigma'}-x_\sigma|$,
which can be a substantial fraction of the full occupation range
as shown in Figs.~\ref{fig_3}(d)--\ref{fig_3}(f).
In the PM solution, where all flavors are equivalent,
such flavor exchanges have a high acceptance ratio
[Fig.~\ref{fig_2}(c)].
Thus, the combination of large occupation changes and a high acceptance ratio
likely contributes to the small integrated autocorrelation time in the PM solution.
The other double-flip updates mainly produce much smaller $|\Delta x_{\sigma}|$ 
and have lower acceptance ratios than the type-1 update [Fig.~\ref{fig_2}(c)].
They therefore appear to contribute less to the improvement of the sampling of the occupation.
\section{Conclusion}
We investigated the sampling efficiency of several Monte Carlo update schemes 
for the doped SU($3$) Fermi-Hubbard model at low temperatures in the strong-coupling regime.
We found that the simple $+$ double-flip update scheme substantially reduces the integrated autocorrelation time
of the majority-flavor occupation in the FM solution
despite the relatively low acceptance ratio of the double-flip update.
Analysis of the configuration changes caused by accepted updates
shows that the type-1 double-flip update transfers existing segments between flavors 
and produces occupation changes on the characteristic scale of the vacancy of the FM majority flavor. 
These results highlight the importance of examining the magnitude of configuration changes
caused by accepted updates when evaluating Monte Carlo sampling efficiency.

\section*{Acknowledgement}
Parts of the numerical calculations were performed 
on the supercomputing systems at ISSP, the University of Tokyo. 
The ALPS libraries~\cite{BauerEtAl2011} were used for some of the simulations.
This work was supported by Grant-in-Aid 
for Scientific Research from JSPS, KAKENHI Grants 
No. JP25K17327 (to K. Y.), No. JP22K03525, No. JP25H01521, and
No. JP25H01398 (to A. K.).
J. F. was supported by the Science Tokyo Support Program for Doctoral Students, 
funded by the Universities for International Research Excellence. 
K. Y. was also supported by JSPS Program for Forming Japan's Peak Research Universities (J-PEAKS) Grant No. JPJS00420230008,
Hirose Foundation, Fujikura Foundation, Toyota Riken Scholar Program, and Support Center for Advanced Telecommunications Technology Research.

\end{document}